\documentclass{PoS}
\usepackage{psfig,epsfig,colordvi,amssymb,amsmath}

\def\lsim{\mathrel{\rlap{\lower4pt\hbox{\hskip1pt$\sim$}}
    \raise1pt\hbox{$<$}}}                
\def\gsim{\mathrel{\rlap{\lower4pt\hbox{\hskip1pt$\sim$}}
    \raise1pt\hbox{$>$}}}                

\newcommand{\be}{\begin{equation}}
\newcommand{\ee}{\end{equation}}
\newcommand{\bea}{\begin{eqnarray}} 
\newcommand{\eea}{\end{eqnarray}}

 \newcommand{\pslash}{{\not{\hspace{-0.08cm}p}}}  
 \newcommand{\ve}{\varepsilon}  
 \newcommand{\csw}{\, c_{\rm SW}}

\title{${\cal O}(a^2)$ corrections to the propagator and bilinears of
  Wilson / clover fermions}

\ShortTitle{${\cal O}(a^2)$ corrections to the fermion propagator and
  bilinears $\overline\Psi\Gamma\Psi$}

\author{\speaker{Martha Constantinou}\\
        Department of Physics, University of Cyprus,\\
P.O.Box 20537, Nicosia CY-1678, Cyprus\\
        E-mail: \email{phpgmc1@ucy.ac.cy}}

\author{Haralambos Panagopoulos\\
        Department of Physics, University of Cyprus,\\
P.O.Box 20537, Nicosia CY-1678, Cyprus\\
        E-mail: \email{haris@ucy.ac.cy}}

\author{Fotos Stylianou\\
        Department of Physics, University of Cyprus,\\
P.O.Box 20537, Nicosia CY-1678, Cyprus\\
        E-mail: \email{fotosstylianou@yahoo.com}}

\abstract{We present the corrections to the fermion propagator, to
  \underline{second order} in the lattice spacing $a$, in 1-loop
  perturbation theory. The fermions are described by the clover action
  and for the gluons we use a 3-parameter family of Symanzik improved
  actions. Our calculation has been carried out in a general covariant
  gauge. The results are provided as a polynomial of the clover
  parameter $c_{\rm SW}$, and are tabulated for 10 popular sets of the
  Symanzik coefficients (Plaquette, Tree-level Symanzik, Iwasaki, TILW
  and DBW2 action).

We also study the $O(a^2)$ corrections to matrix elements of fermion bilinear
operators that have the form $\overline\Psi\Gamma\Psi$, where $\Gamma$
denotes all possible distinct products of Dirac matrices. These correction terms 
are essential ingredients for improving, to $O(a^2)$, the matrix elements of the
fermion operators.

Our results are applicable also to the case of twisted mass
fermions.

\medskip
\noindent
A longer write-up of this work, including non-perturbative results, is
in preparation together with V. Gim\'enez, V. Lubicz and D. Palao
\cite{CGLPPS}. }

\FullConference{The XXVI International Symposium on Lattice Field Theory\\
		 July 14-19, 2008\\
		 Williamsburg, Virginia, USA}

\begin{document}

\section{Introduction}

Over the years, many efforts have been made for ${\cal O}(a^1)$ 
improvement in lattice observables, which in many cases is automatic
by virtue of symmetries of the fermion action. According to Symanzik's
program \cite{Symanzik}, one can improve the action by adding irrelevant
operators. Also, in the twisted mass formulation of QCD~\cite{FGSW}
at maximal twist, certain observables are ${\cal O}(a^1)$ improved, by
symmetry considerations.  

So far, in the literature there appear two kinds of perturbative
evaluations pertaining to the fermion propagator and bilinears of the
form $\overline\Psi\Gamma\Psi$ ($\Gamma$ denotes all possible distinct
products of Dirac matrices). On the one hand, there are 1-loop
computations for ${\cal O}(a^1)$ corrections, with an arbitrary fermion
mass~\cite{ANTU,CGHPRSS}. On the other hand, there are 2-loop
calculations at ${\cal O}(a^0)$ level, for massless fermions~\cite{SP}. 
1-loop computations of ${\cal O}(a^2)$ corrections did not exist to
date; indeed they present some novel difficulties. In particular,
extending ${\cal O}(a^0)$ calculations up to ${\cal O}(a^1)$ does not
bring in any novel types of singularities. For instance, terms which
were convergent to ${\cal O}(a^0)$ may now develop an infrared (IR)
logarithmic singularity at worst in 4 dimensions and the way to treat
such singularities is well known. In most of the cases, e.g. for
$m=0$, terms which were already IR divergent to ${\cal O}(a^0)$
will not contribute to ${\cal O}(a^1)$, by parity of loop integration. 
On the contrary, the IR singularities encountered at ${\cal
  O}(a^2)$ are present even in 6 dimensions, making their extraction
more delicate.

\section{Description of the calculation}
\label{method}
Our calculation is performed for clover fermions, keeping the
coefficient $\csw$ as a free parameter. The action describing $N_f$
flavors of degenerate clover (SW) fermions is given in Ref.~\cite{SP}. 
We work with massless fermions ($m_{\rm o}=0$), which simplifies the
algebraic expressions, but at the same time requires special treatment
for the IR singularities. By taking $m_{\rm o}=0$, our calculation
and results are identical also for the twisted mass action in the
chiral limit.

For the gluon part we employ the Symanzik improved action, involving
Wilson loops with 4 and 6 links; for the Symanzik coefficients,
$c_i$, multiplying each Wilson loop, we choose 10 sets of values that
are widely used in numerical simulations; these are tabulated in
Ref.~\cite{CGLPPS}.

\medskip
The Feynman diagrams that enter this computation are shown in
Fig. 1; diagrams 1 and 2 contribute to the fermion propagator, while
diagram 3 is relevant to the bilinears' improvement.

\begin{center}
\psfig{figure=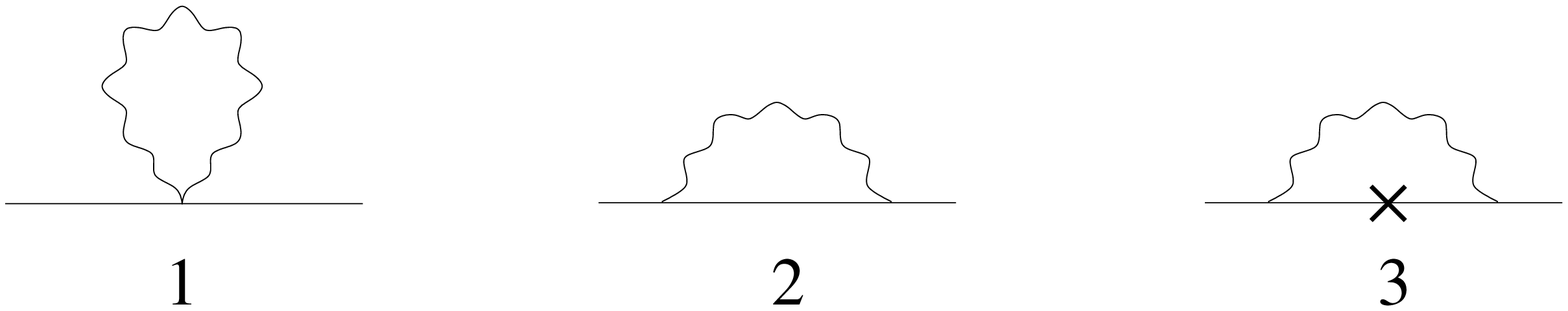,height=1.75truecm}
\end{center}
{\small 
\begin{center}
\begin{minipage}{13.5cm}
{\bf Fig. 1:} Diagrams contributing to the improvement of the
propagator (1, 2) and the bilinears (3). A wavy (solid) line represents
gluons (fermions). A cross denotes an insertion of $\Gamma$
(Eq.~(\ref{Gamma})). 
\end{minipage}
\end{center}}
\vspace{0.5cm}

For the algebraic operations involved in manipulating  lattice Feynman
diagrams, we make use of our symbolic package in Mathematica. Next, we
briefly describe the required steps:

\vskip 0.10cm \noindent $\bullet$
The evaluation of each diagram starts with the contraction among
vertices, which is performed automatically once the vertices and the
topology of the diagram are specified. The outcome of the \newpage 
\noindent contraction is a preliminary expression for the diagram
under study; there follow simplifications of the color dependence,
$\rm Dirac$ matrices and tensor structures. We also fully exploit
symmetries of the theory to limit the proliferation of the algebraic
expressions. 

\vskip 0.10cm \noindent $\bullet$
The above simplifications are followed by the extraction of all
functional dependencies on the external momentum $p$ (divergent,
convergent terms) and the lattice spacing (terms of order
$a^0,\,a^1,\,a^2$). The convergent terms can be treated by naive
Taylor expansion in $a\,p$ to the desired order. On the contrary, the
isolation of the logarithms and non-Lorentz invariant terms is
achieved as described below. As a first task we want to  reduce the
number of infrared divergent integrals to a minimal set. To do this, we use
{\it{iteratively}} two kinds of subtractions for the propagator, so that all
primitively divergent integrals (initially depending on the fermion
and the Symanzik propagator) are expressed in terms of the Wilson
propagator. The subtraction for the gluon propagator reads
\be
D(q) =  D_{plaq}(q) + D_{plaq}(q) \,\big(D^{-1}_{plaq}(q) - D^{-1}(q)\big)\, D(q)
\label{sub2}
\ee
\be
D^{\mu\nu}_{plaq}(q) = \frac{\delta_{\mu\nu}}{\hat{q}^2} -
(1-\lambda)\frac{\hat{q}_\mu\,\hat{q}_\nu}{(\hat{q}^2)^2},\quad
\hat{q}_\mu = 2 \sin(\frac{q_\mu}{2}),\quad
\hat{q}^2 =\sum_\mu \hat{q}_\mu^2
\ee
where $D^{\mu\nu}$ is the $4\times4$ Symanzik propagator. The
matrix $( D^{-1}_{plaq}(q) - D^{-1}(q) )$, which is
${\cal O}(q^4)$, is independent of the gauge parameter, $\lambda$, and
can be obtained in closed form, as a polynomial in $\hat{q}_\mu$.

\vskip 0.10cm \noindent $\bullet$
The most laborious part is the computation of the divergent terms,
which is performed in a noninteger number of dimensions $D>4$.
Ultraviolet divergences are explicitly isolated \`a la Zimmermann and
evaluated as in the continuum. The remainders are $D$-dimensional,
parameter-free, zero momentum lattice integrals which can be recast in
terms of Bessel functions, and finally expressed as sums of a pole
part plus numerical constants. We analytically evaluated an extensive
basis of superficially divergent loop integrals, which is presented in
Ref.~\cite{CGLPPS}. A few of these integrals are very demanding,
because they must be evaluated to two further orders in $a$, beyond
the order at which an IR divergence initially sets in. As a
consequence, their evaluation requires going to $D>6$ dimensions, with
due care to take into account all possible sources of ${\cal O}(a^2)$
corrections. These 
integrals form a sufficient basis for all integrals which can appear in
any ${\cal O}(a^2)$ 1-loop calculation.

\vskip 0.10cm \noindent $\bullet$
The required numerical integrations over loop momenta are performed by
highly optimized Fortran programs; these are generated by our
Mathematica `integrator' routine. Each integral is expressed as a sum
over the discrete Brillouin zone of finite lattices, with varying size
$L$ ($L^4 \leq 128^4$), and evaluated for all values of the
Symanzik coefficients which we considered.

\vskip 0.10cm \noindent $\bullet$
The last part of the evaluation is the extrapolation of the numerical
results to infinite lattice size. This procedure entails a systematic
error, which is reliably estimated, using a sophisticated inference
technique; for 1-loop quantities we expect a fractional error smaller
than $10^{-8}$.

\section{Correction to the fermion propagator}
\label{prop}

The 1-loop corrections to the fermion propagator arise from the
evaluation of diagrams 1 and 2 in Fig. 1. Capitani et al. 
\cite{CGHPRSS} have calculated the first order terms in the lattice
spacing for massive Wilson fermions and the Plaquette action for
gluons. We carried out this calculation beyond the first order
correction, taking into account all terms up to ${\cal O}(a^2)$ and
considering a general Symanzik improved gluon action. Our results, to
${\cal O}(a^1)$, are in perfect agreement with those of Ref.~\cite{CGHPRSS}. 

The following equation is the  total expression for the inverse
propagator $S^{-1}$ as a function of the external momentum $p$, the
coupling constant $g$, the number of colors $N$, the clover
coefficient $c_{\rm SW},$ and the gauge parameter $\lambda$. The
quantities $\ve^{(i,j)}$ appearing in our results for $S^{-1}$ are
numerical coefficients depending on the Symanzik parameters,
calculated for each action we have considered; they are tabulated in
Ref. \cite{CGLPPS}. In Eq.~(\ref{ve}) and Eqs. (\ref{ve_SP}),
(\ref{ve_VA}) we present the values of $\ve^{(i,j)}$ for 
the Plaquette and Iwasaki actions (top and bottom numbers,
respectively); only 5 decimal points are shown, due to lack of space
\bea
\hspace{-0.85cm}
S^{-1}(p) &=& i \, \pslash+ \frac{a}{2} p^2 - i \frac{a^2}{6}\,\pslash^3
- i\,\pslash \, \tilde{g}^2 \Big[ \,
      \ve^{(0,1)} -4.79201\,\lambda  +\ve^{(0,2)}\csw +\ve^{(0,3)}\csw^2
   + \lambda \ln(a^2 p^2) \Big]  \nonumber \\[1.5ex]
&-& a\, p^2\, \tilde{g}^2 \Big[
     \ve^{(1,1)} -3.86388\,\lambda  +\ve^{(1,2)}\csw +\ve^{(1,3)}\csw^2
 - \frac{1}{2} \left( 3 - 2 \,\lambda  -  3 \csw \right) \, \ln(a^2 p^2)
        \Big] \nonumber \\ [1.5ex]
&-& i\,a^2\, \pslash^3 \, \tilde{g}^2 \Big[\,
\ve^{(2,1)} +0.50700\,\lambda +\ve^{(2,2)}\csw + \ve^{(2,3)}\csw^2
+\left(\frac{101}{120}-\frac{11}{30}\,C_2
-\frac{\lambda}{6} \right)\,\ln(a^2 p^2)\Big]\nonumber \\[1.5ex]
&-& i\,a^2\, p^2\,\pslash \, \tilde{g}^2 \Big[\,
\ve^{(2,4)} +1.51605\,\lambda +\ve^{(2,5)}\csw + \ve^{(2,6)}\csw^2 \nonumber\\[1.5ex]
&&\phantom{i\,a^2\, p^2\,\pslash \, \tilde{g}^2}
+\left(\frac{59}{240}+\frac{c_1}{2}+\frac{C_2}{60}
-\frac{1}{4}\left(\frac{3}{2}\lambda+
\csw+\csw^2\right)\right) \, \ln(a^2 p^2)\Big] \nonumber \\ [1.5ex]
&-& i\,a^2\,\pslash\, \frac{\sum_\mu p^4_\mu}{p^2}\,\tilde{g}^2\Big[\,
-\frac{3}{80}-\frac{C_2}{10} -\frac{5}{48}\lambda \Big] 
\label{propagator}
\eea
We define $\tilde{g}^2 \equiv g^2 C_F/(16\pi^2)$, $C_F=(N^2-1)/(2\,N)$, 
$C_2=c_1 - c_2 - c_3$ and $\pslash^3= \sum_\mu \gamma_\mu p_\mu^3$\,; 
the specific values $\lambda=1\,(\lambda=0)$ correspond to the Feynman 
(Landau) gauge. We observe that the ${\cal O}(a^1)$ logarithms are 
independent of the Symanzik coefficients $c_i$\,; on the contrary ${\cal O}(a^2)$ 
logarithms have a mild dependence on $c_i$\,.
\bea
\begin{array}{rclrclrclrclrcl}
\hspace{-0.25cm}
\displaystyle
 \ve^{(0,1)}&=&{\,\,16.64441\atop\,\,\phantom{1}8.11657},\quad &
 \ve^{(0,2)}&=&{-2.24887\atop-1.60101},\quad &
 \ve^{(0,3)}&=&{-1.39727\atop-0.97321},\quad &
 \ve^{(1,1)}&=&{\,\,12.82693\atop\,\,\phantom{1}7.40724}\nonumber\\[2.25ex]
\hspace{-0.25cm}
 \ve^{(1,2)}&=&{-5.20234\atop-3.88884},\quad &
 \ve^{(1,3)}&=&{-0.08173\atop-0.06103},\quad &
 \ve^{(2,1)}&=&{-4.74536\atop-3.20180},\quad &
 \ve^{(2,2)}&=&{\phantom{-}0.02029\atop\phantom{-}0.08250}\nonumber\\[2.25ex]
\hspace{-0.25cm}
 \ve^{(2,3)}&=&{\phantom{-}0.10349\atop\phantom{-}0.04192},\quad &
  \ve^{(2,4)}&=&{-1.50481\atop-0.62022},\quad &
 \ve^{(2,5)}&=&{\phantom{-}0.70358\atop\phantom{-}0.55587},\quad &
 \ve^{(2,6)}&=&{\phantom{-}0.53432\atop\phantom{-}0.41846} 
 \end{array}
\label{ve}\\[-3ex]
\eea

\section{Improved operators}
\label{oper}

In the context of this work we also compute the contributions up to ${\cal O}(a^2)$
to the forward matrix elements of local fermion operators
that have the form $\overline\Psi\Gamma\Psi$. $\Gamma$ corresponds to the
following set of products of the Dirac matrices 
\be
\Gamma = \hat{1}\,\,({\footnotesize{\rm scalar}}),\quad 
\gamma^5\,\,({\footnotesize{\rm pseudoscalar}}),\quad
\gamma_\mu\,\,({\footnotesize{\rm vector}}),\quad
\gamma^5\gamma_\mu\,\,({\footnotesize{\rm axial}}),\quad
\frac{1}{2}\gamma^5[\gamma_\mu,\gamma_\nu]\,\,({\footnotesize{\rm tensor}})
\label{Gamma}
\ee
The ${\cal O}(a^2)$ correction terms are derived from the evaluation
of diagram 3 shown in Fig. 1. One may improve the local bilinears by
the addition of higher-dimension operators
\begin{equation} 
 \left({\cal O}^\Gamma \right)^{\rm imp} =
 \overline\Psi \Gamma \Psi + a\, g^2_0\,\sum_{i=1}^n \,k^\Gamma_i \,\overline\Psi \,Q^\Gamma_i \,\Psi
+ a^2 \, g^2_0\,\sum_{i=1}^{\tilde n} \,\tilde{k}^\Gamma_i \,\overline\Psi\, \tilde{Q}^\Gamma_i  \,\Psi
 \label{Gamimp0}
 \end{equation}
where the first term is the unimproved operator and $\Psi Q^\Gamma_i \Psi$
($\Psi  \tilde{Q}^\Gamma_i \Psi$) are operators with the same symmetries as the
original ones, but with dimension higher by one (two) units. To achieve ${\cal O}(a^2)$ 
improvement, we must choose $k^\Gamma_i$ and $\tilde{k}^\Gamma_i$ appropriately, in order to
cancel out all ${\cal O}(a^1,\,a^2)$ contributions in matrix elements.

In the rest of this section we show our results for the 1-loop
corrections to the amputated 2-point Green's function (diagram 3 of
Fig. 1), at momentum $p$\,: 
$\Lambda^\Gamma(p) = \langle\Psi\,\left(\overline\Psi\Gamma\Psi\right)\,\overline\Psi\rangle_{(p)}^{amp}$.
The values of all the Symanzik dependent coefficients,
$\ve_S,\,\ve_P,\,\ve_V,\,\ve_A,\,\ve_T$, with their systematic errors,
can be found in Ref.~\cite{CGLPPS}.
We begin with the ${\cal O}(a^2)$ corrected expressions for
$\Lambda^S(p)$ and $\Lambda^P(p)$; including the tree-level term, we obtain
\noindent
\bea\hspace{-.5cm}
\Lambda^S(p)  &=& 1+ 
\tilde{g}^2\Big[ \,\ve_S^{(0,1)} +5.79201\,\lambda  +\ve_S^{(0,2)}\csw +\ve_S^{(0,3)}\csw^2
   -\ln(a^2 p^2)\left(3+\lambda\right)\Big]\nonumber \\[1.25ex]
&+&\,a\,i\,\pslash\, \tilde{g}^2 \Big[ \,
\ve_S^{(1,1)} -3.93576\,\lambda +\ve_S^{(1,2)}\,\csw+\ve_S^{(1,3)}\,\csw^2
+\left(\frac{3}{2}+\lambda +\frac{3}{2}\csw \right)\ln(a^2 p^2)\Big]\nonumber \\[1.25ex]
&+&\,a^2\, p^2 \,  \tilde{g}^2 \Big[\,
\ve_S^{(2,1)} -2.27359\,\lambda +\ve_S^{(2,2)}\csw + \ve_S^{(2,3)}\csw^2
+\hspace{-0.01cm}\left(-\frac{1}{4}+\frac{3}{4}\lambda +\frac{3}{2}\csw \right)\ln(a^2 p^2)\Big]
\nonumber \\[1.25ex]
&+&\,a^2\,\frac{\sum_\mu p_\mu^4}{p^2}\,  \tilde{g}^2 
\Big[\,\frac{13}{24}+\frac{C_2}{2} -\frac{\lambda}{8}\Bigr]
\label{scalar}\\[3ex]
\Lambda^P(p)  &=& \gamma^5 + \gamma^5\, \tilde{g}^2 \Big[ \,
      \ve_P^{(0,1)} +5.79201\,\lambda  +\ve_P^{(0,2)}\csw^2
   -\ln(a^2 p^2)\left(3+\lambda\right)\Big]\nonumber \\[1.25ex]
&+&\,a^2\, p^2 \,\gamma^5\, \tilde{g}^2 \Big[\,
\ve_P^{(2,1)} -0.83810\,\lambda +\ve_P^{(2,2)}\csw^2
+\hspace{-0.01cm}\left(-\frac{1}{4}+\frac{1}{4}\lambda\right)\ln(a^2 p^2)\Big]
\nonumber \\[1.25ex]
&+&\,a^2\, \frac{\sum_\mu p_\mu^4}{p^2}\,\gamma^5\, \tilde{g}^2 
\Big[\,\frac{13}{24}+\frac{C_2}{2} -\frac{\lambda}{8} \Bigr]
\label{pseudoscalar}
\eea
$\Lambda^P(p)$ is free of ${\cal O}(a^1)$ terms and all contributions
linear in $\csw$ vanish. The values of the numerical coefficients
$\ve_S$ and $\ve_P$ for the Wilson and Iwasaki gluon action are
\bea
\begin{array}{rclrclrclrclrcl}
\hspace{-0.25cm}
 \ve_S^{(0,1)}&=&{\phantom{-}0.30800\atop\phantom{-}0.74092},\,\,&
 \ve_S^{(0,2)}&=&{\phantom{-}9.98678\atop\phantom{-}6.90168},\,\,&
 \ve_S^{(0,3)}&=&{\phantom{-}0.01689\atop-0.29335},\,\,&
 \ve_S^{(1,1)}&=&{\phantom{-}0.65863\atop-0.05097},\,\,&
 \ve_S^{(1,2)}&=&{-4.20299\atop-2.88571} \nonumber\\[2.25ex]
\hspace{-0.25cm}
 \ve_S^{(1,3)}&=&{-1.28605\atop-0.90950},\,\,&
 \ve_S^{(2,1)}&=&{\phantom{-}2.60041\atop\phantom{-}2.02123},\,\,&
 \ve_S^{(2,2)}&=&{-4.15080\atop-3.23460},\,\,&
 \ve_S^{(2,3)}&=&{\phantom{-}0.17641\atop\phantom{-}0.23450},\,\,&
 \ve_P^{(0,1)}&=&{\phantom{-}9.95103\atop\phantom{-}6.55611} \nonumber\\[2.25ex]
\hspace{-0.25cm}
 \ve_P^{(0,2)}&=&{\phantom{-}3.43328\atop\phantom{-}2.25383},\,\,&
 \ve_P^{(2,1)}&=&{\phantom{-}0.84420\atop\phantom{-}0.66991},\,\,& 
 \ve_P^{(2,2)}&=&{-0.25823\atop-0.30221} 
\end{array}
\label{ve_SP}\\[-3ex]
\eea

The ${\cal O}(a^2)$ corrected expressions for $\Lambda^V(p)$,
$\Lambda^A(p)$ and $\Lambda^T(p)$ are very complicated, in the sense
that there is a variety of momentum contributions and therefore many
Symanzik dependent coefficients, as can be seen from
Eqs. (\ref{vector}) - (\ref{axial}). In fact, we relegate our result
for $\Lambda^T(p)$ to the longer write up~\cite{CGLPPS}. We also list
the coefficients $\ve_V$ and $\ve_A$ for the Wilson and Iwasaki
actions in Eq.~(\ref{ve_VA}).
\vspace{-0.5cm}
\hspace{-0.25cm}
\bea
\Lambda^V(p) &=& \gamma_\mu 
+\frac{\pslash\,p_\mu}{p^2}\, \tilde{g}^2 \Big[-2\,\lambda \Big]
+ \gamma_\mu \,\tilde{g}^2 \Big[ \,\ve_V^{(0,1)} +4.79201\,\lambda
+\ve_V^{(0,2)}\csw+\ve_V^{(0,3)}\csw^2 -\lambda\ln(a^2 p^2)\Big]\nonumber \\[1.25ex]
&+&a\,i\,p_\mu \,\tilde{g}^2 \Big[\,\ve_V^{(1,1)} -0.93576\,\lambda
+\ve_V^{(1,2)}\csw+\ve_V^{(1,3)}\csw^2
+\left(-3+\lambda+3\,\csw
\right)\ln(a^2 p^2) \Big]\nonumber \\[1.25ex] 
&+&a^2\,\gamma_\mu\, p_\mu^2 \,\tilde{g}^2 \Big[\,\ve_V^{(2,1)}+\frac{\lambda}{8}
+\ve_V^{(2,2)}\csw+\ve_V^{(2,3)}\csw^2 
+\left(-\frac{53}{120}+\frac{11}{10}\,C_2 \right)
\ln(a^2 p^2)\Big]\nonumber \\[1.25ex] 
&+&a^2\,\gamma_\mu\, p^2 \, \tilde{g}^2 \Big[\,\ve_V^{(2,4)} -0.81104\,\lambda
+\ve_V^{(2,5)}\csw+\ve_V^{(2,6)}\csw^2\nonumber \\[1.25ex] 
&&\phantom{a^2\,\gamma_\mu\, p^2\, \tilde{g}^2 \Big[\,}
+\left(\frac{11}{240}-\frac{c_1}{2}-\frac{C_2}{60}
+\frac{\lambda}{8}-\frac{5}{12}\csw+\frac{\csw^2}{4}\right)\ln(a^2
p^2)\Big]\nonumber\\[1.25ex] 
&+&a^2\,\pslash\, p_\mu\, \tilde{g}^2 \Big[\,\ve_V^{(2,7)} +0.24364\,\lambda
+\ve_V^{(2,8)}\csw+\ve_V^{(2,9)}\csw^2\nonumber \\[1.25ex] 
&&\phantom{a^2\, \pslash\,p_\mu\, \tilde{g}^2 \Big[\,}
+\left(-\frac{149}{120}-c_1-\frac{C_2}{30}
+\frac{\lambda}{4}+\frac{\csw}{6}+\frac{\csw^2}{2}\right)\ln(a^2
p^2)\Big]\nonumber \\[1.25ex] 
&+&a^2\,\gamma_\mu\,\frac{\sum_\rho p_\rho^4}{p^2}\,\tilde{g}^2 \Big[\,
\frac{3}{80}+\frac{C_2}{10}+\frac{5}{48}\,\lambda \Bigr] 
+a^2\,\frac{\pslash^3\,p_\mu}{p^2}\, \tilde{g}^2 \Big[
-\frac{101}{60}+\frac{11}{15}\,C_2 +\frac{\lambda}{3} \Bigr] \nonumber \\[1.25ex] 
&+&a^2\,\frac{\pslash\, p_\mu^3}{p^2}\, \tilde{g}^2 \Big[
-\frac{1}{60}+\frac{2}{5}\,C_2 +\frac{\lambda}{12} \Bigr] 
+a^2\,\frac{\pslash\,p_\mu\,\sum_\rho
p_\rho^4}{(p^2)^2}\, \tilde{g}^2 \Big[
-\frac{3}{40}-\frac{C_2}{5} -\frac{5}{24}\,\lambda \Bigr]
\label{vector}
\\[2.5ex] 
\Lambda^A(p)&=& \gamma^5\,\gamma_\mu
+\frac{\gamma^5\,\pslash\,p_\mu}{p^2}\, \, \tilde{g}^2 \Big[-2\,\lambda \Big]\nonumber \\[1.25ex]
&+&\gamma^5\,\gamma_\mu \, \tilde{g}^2 \Big[ \,\ve_A^{(0,1)} +4.79201\,\lambda
+\ve_A^{(0,2)}\csw+\ve_A^{(0,3)}\csw^2 -\lambda\ln(a^2 p^2)\Big]\nonumber \\[1.25ex]
&+&a\,i\,\gamma^5\,\left(\gamma_\mu\, \pslash-p_\mu\right) \, \tilde{g}^2 
\Big[\,\ve_A^{(1,1)} -2.93576\,\lambda
+\ve_A^{(1,2)}\csw+\ve_A^{(1,3)}\csw^2 +\lambda\ln(a^2 p^2) \Big]\nonumber \\[1.25ex] 
&+&a^2\,\gamma^5\,\gamma_\mu\, p_\mu^2 \, \tilde{g}^2 \Big[\,\ve_A^{(2,1)}+\frac{\lambda}{8}
+\ve_A^{(2,2)}\csw+\ve_A^{(2,3)}\csw^2 +
\left(-\frac{53}{120}+\frac{11}{10}\,C_2 \right)\ln(a^2 p^2)\Big]\nonumber \\[1.25ex] 
&+&a^2\,\gamma^5\,\gamma_\mu\, p^2 \, \tilde{g}^2 \Big[\,\ve_A^{(2,4)} -1.74652\,\lambda
+\ve_A^{(2,5)}\csw+\ve_A^{(2,6)}\csw^2\nonumber \\[1.25ex] 
&&\phantom{a^2\,\gamma^5\,\gamma_\mu\, p^2\, \tilde{g}^2 \Big[\,}
+\left(-\frac{109}{240}-\frac{c_1}{2}-\frac{C_2}{60}
+\frac{5}{8}\lambda+\frac{7}{12}\csw-\frac{\csw^2}{4}\right)\ln(a^2
p^2)\Big]\nonumber\\[1.25ex] 
&+&a^2\,\gamma^5\,\pslash\, p_\mu\, \tilde{g}^2 \Big[\,\ve_A^{(2,7)} +1.11462\,\lambda
+\ve_A^{(2,8)}\csw+\ve_A^{(2,9)}\csw^2\nonumber \\[1.25ex] 
&&\phantom{a^2\,\gamma^5\,\pslash\,p_\mu\, \tilde{g}^2 \Big[\,}
+\left(\frac{91}{120}-c_1-\frac{C_2}{30}
-\frac{3}{4}\lambda-\frac{5}{6}\csw-\frac{\csw^2}{2}\right)\ln(a^2
p^2)\Big]\nonumber \\[1.25ex] 
&+&a^2\,\gamma^5\,\gamma_\mu\,\frac{\sum_\rho p_\rho^4}{p^2}\,\tilde{g}^2 \Big[\,
\frac{3}{80}+\frac{C_2}{10}+\frac{5}{48}\,\lambda \Bigr] 
+a^2\,\gamma^5\frac{\pslash^3\,p_\mu}{p^2}\, \tilde{g}^2 \Big[
-\frac{101}{60}+\frac{11}{15}\,C_2 +\frac{\lambda}{3} \Bigr] \nonumber \\[1.25ex] 
&+&a^2\,\gamma^5\frac{\pslash\, p_\mu^3}{p^2}\, \tilde{g}^2 \Big[
-\frac{1}{60}+\frac{2}{5}\,C_2 +\frac{\lambda}{12} \Bigr] 
+a^2\,\gamma^5\frac{\pslash\,p_\mu\,\sum_\rho
p_\rho^4}{(p^2)^2}\, \tilde{g}^2 \Big[
-\frac{3}{40}-\frac{C_2}{5} -\frac{5}{24}\,\lambda \Bigr]
\label{axial}
\eea

\bea
\begin{array}{rclrclrclrclrcl}
\hspace{-0.25cm}
  \ve_V^{(0,1)}&=&{\phantom{-}3.97338\atop\phantom{-}2.98283},\,\,&
  \ve_V^{(0,2)}&=&{-2.49670\atop-1.72542},\,\,&
  \ve_V^{(0,3)}&=&{\phantom{-}0.85410\atop\phantom{-}0.63680},\,\,&
  \ve_V^{(1,1)}&=&{\phantom{-}2.71098\atop\phantom{-}0.90743},\,\,&
  \ve_V^{(1,2)}&=&{-1.84814\atop-0.80352} \nonumber\\[2.15ex]
\hspace{-0.15cm}
  \ve_V^{(1,3)}&=&{-0.39053\atop-0.35601},\,\,&
  \ve_V^{(2,1)}&=&{\phantom{-}1.55410\atop\phantom{-}1.45731},\,\,&
  \ve_V^{(2,2)}&=&{\phantom{-}0.32907\atop\phantom{-}0.08590},\,\,&
  \ve_V^{(2,3)}&=&{-0.00602\atop\phantom{-}0.07935},\,\,&
  \ve_V^{(2,4)}&=&{\phantom{-}0.25007\atop-0.26685}\nonumber\\[2.15ex]
\hspace{-0.15cm}
  \ve_V^{(2,5)}&=&{\phantom{-}0.88599\atop\phantom{-}0.71279},\,\,&
  \ve_V^{(2,6)}&=&{-0.30036\atop-0.25078},\,\,&
  \ve_V^{(2,7)}&=&{\phantom{-}1.27888\atop\phantom{-}0.76263},\,\,&
  \ve_V^{(2,8)}&=&{\phantom{-}0.27776\atop\phantom{-}0.29755},\,\,&
  \ve_V^{(2,9)}&=&{-0.35475\atop-0.18427}\nonumber\\[2.15ex]
\hspace{-0.15cm}
  \ve_A^{(0,1)}&=&{-0.84813\atop\phantom{-}0.07524},\,\,&
  \ve_A^{(0,2)}&=&{\phantom{-}2.49670\atop\phantom{-}1.72542},\,\,&
  \ve_A^{(0,3)}&=&{-0.85410\atop-0.63680},\,\,&
  \ve_A^{(1,1)}&=&{\phantom{-}1.34275\atop\phantom{-}0.26850},\,\,&
  \ve_A^{(1,2)}&=&{-1.71809\atop-1.23802}\nonumber\\[2.15ex]
\hspace{-0.15cm}
  \ve_A^{(1,3)}&=&{\phantom{-}0.13018\atop\phantom{-}0.11867},\,\,&
  \ve_A^{(2,1)}&=&{\phantom{-}0.38791\atop\phantom{-}0.05918},\,\,&
  \ve_A^{(2,2)}&=&{\phantom{-}1.85117\atop\phantom{-}1.57070},\,\,&
  \ve_A^{(2,3)}&=&{-0.09309\atop-0.13933},\,\,&
  \ve_A^{(2,4)}&=&{\phantom{-}1.63504\atop\phantom{-}0.68458}\nonumber\\[2.15ex]
\hspace{-0.15cm}
  \ve_A^{(2,5)}&=&{-1.59946\atop-1.24801},\,\,&
  \ve_A^{(2,6)}&=&{\phantom{-}0.33390\atop\phantom{-}0.26827},\,\,&
  \ve_A^{(2,7)}&=&{\phantom{-}0.41759\atop\phantom{-}1.05772},\,\,&
  \ve_A^{(2,8)}&=&{\phantom{-}0.39585\atop\phantom{-}0.18672},\,\,&
  \ve_A^{(2,9)}&=&{\phantom{-}0.31972\atop\phantom{-}0.17429} 
\end{array}
\\ [0.5ex] \label{ve_VA}
\eea

\section{Conclusions}

Our results show clearly that ${\cal O}(a^2)$ effects are quite
pronounced in the Green's functions we have considered. The ${\cal
  O}(a^2)$ contributions which we have calculated can be directly used
in order to construct improved operators, bringing the chiral limit
within reach. Possible follow-ups to the present work include:
\vspace{-0.25cm}
\begin{itemize}
\item Extending to the case of nonzero renormalized mass.
\vspace{-0.25cm}
\item Improvement of higher-dimension bilinear operators,
such as those involved in hadronic form factors, and of 4-fermi
operators.
\end{itemize}
\vspace{-0.25cm}
A comparison with non-perturbative estimates of matrix elements,
coming from numerical simulations, will be presented in Ref.~\cite{CGLPPS}.


\end{document}